\documentclass[journal=jacsat,manuscript=article]{achemso}

\usepackage{chemformula} 
\usepackage[T1]{fontenc} 
\usepackage[version=3]{mhchem}
\author{Zheng Sun}
\email{zhengsun@pitt.edu}
\altaffiliation{Contributed equally to this work}
\author{Jonathan Beaumariage}
\altaffiliation{Contributed equally to this work}
\affiliation[University of Pittsburgh]
{Department of Physics and Astronomy, University of Pittsburgh, Pittsburgh, PA 15260, USA}

\author{Qingrui Cao}
\affiliation[Carnegie Mellon University]
{Department of Physics, Carnegie Mellon University 15213, USA}

\author{Kenji Watanabe}
\author{Takashi Taniguchi}
\affiliation[National Institute for Materials Science]
{National Institute for Materials Science, Tsukuba, Ibaraki 305-0044, Japan}

\author{Benjamin Matthew Hunt}
\affiliation[Carnegie Mellon University]
{Department of Physics, Carnegie Mellon University 15213, USA}

\author{David Snoke}
\affiliation[University of Pittsburgh]
{Department of Physics and Astronomy, University of Pittsburgh, Pittsburgh, PA 15260, USA}
\email{snoke@pitt.edu}
\phone{+14126249007}
\fax{+14126249163}
\title{Observation of the Interlayer Exciton Gases in WSe$_2$ -p:WSe$_2$ Heterostructures}

\abbreviations{IXs,TMDs,HSs,2D}

\begin{document}


\begin{abstract}
  Interlayer excitons (IXs) possess a much longer lifetime than intralayer excitons due to the spatial separation of the electrons and holes; hence, they have been pursued to create exciton condensates for decades. The recent emergence of two-dimensional (2D) materials, such as transition metal dichalcogenides (TMDs), and of their van der Waals heterostructures (HSs), in which two different 2D materials are layered together, has created new opportunities to study IXs. Here we present the observation of IX gases within two stacked structures consisting of hBN/WSe$_2$/hBN/p:WSe$_2$/hBN. The IX energy of the two different structures differed by 82 meV due to the different thickness of the hBN spacer layer between the TMD layers. We demonstrate that the lifetime of the IXs is shortened when the temperature and the pump power increase. We attribute this nonlinear behavior to an Auger process.  \\
\end{abstract}
\section{Introduction}

Interlayer excitons (IXs) (or spatially indirect excitons) are electrons and holes that are bound by Coulomb interaction but spatially separated in two different quantum wells. For decades III-V and II-IV type quantum wells have been investigated as candidates for IX condensation\cite{Butov2002,snoke2006}. However, the weak binding energy and short lifetime of traditional IXs has hindered that progress. 

Transition metal dichalcogenides (TMDs) with the chemical formula of MX$_2$ (M = Mo, W; X = S, Se) recently have attracted intense attention because of their promising potential for use in the next generation of spin-, valley-, optical and optoelectrical devices\cite{Ross2014a,Koperski2015a,Schaibley2016,Xu2014a,Lopez-Sanchez2013}. Single atomic layer TMDs possess a direct bandgap, extraordinarily strong oscillator strength and larger exciton binding energy than the conventional semiconductors (0.3-0.5 eV).\cite{Chernikov2014,Ugeda2014} Moreover, the various possibilities of assembling stacks of TMDs via weak van der Waals forces makes them an exciting new platform in investigating light-matter interactions\cite{Liu2014c,Sun2017a,Chakraborty2018a}, including microcavity polaritons, using devices in which the TMD layers are embedded in a planar optical cavity. However, optical nonlinearity and the onset of spontaneous coherence in such a system has not yet been achieved, due to the difficulty of fabricating high Q-factor microcavities with a wedge for tunability, resulting in low-density populations and inefficient detunings.\\

On the other hand, a series of studies on hetero-bilayers of WSe$_2$/WS$_2$, MoS$_2$/WS$_2$,  MoS$_2$/WSe$_2$ and MoSe$_2$/WSe$_2$, have been reported in the last few years. Ultrafast charge transfer and the formation of interlayer excitons have been demonstrated in photoexcited TMD heterostructures\cite{Zhou2020a,Schaibley2016,Rivera2015,Hong2014,Rivera2016a,Lee2014,Karni2019,Unuchek2019}. In particular, the IXs within the heterostructures exhibit much longer transition and valley depolarization lifetimes (several hundreds of picoseconds to nanoseconds) compared to conventional semiconductors. The bandgap at the interface of the two different TMDs realigns, leading to type II semiconductors. The fixed vertically aligned IX dipole moment gives rise to the tunability of the IX energy by applying an out-of-plane electric field. However, the electrical tunability is limited by the built-in interfacial electric field which produces pristine homobilayer WSe$_2$/WSe$_2$ structures with negligible built-in field\cite{Wang2018a}. However, a heterostructure with an embedded large-index material such as hexagonal boron nitride (hBN) between the two TMDs has not been studied as thoroughly\cite{Fang2014, Calman2018}. Here in the paper, we look at bilayers of the same TMD material but one of the layers doped, with a spacer. We demonstrate the existence of the IX in a sample with a “thick” hBN spacer, which was previously theoretically predicted \cite{Latini2017}. The structure studied in our paper is a more realistic scenario for future devices fabricated using wide-area growth methods, in which it is desirable to have fewer different element sources.\\

\begin{figure}[!h]
	\centering
	\includegraphics[width=0.9\linewidth]{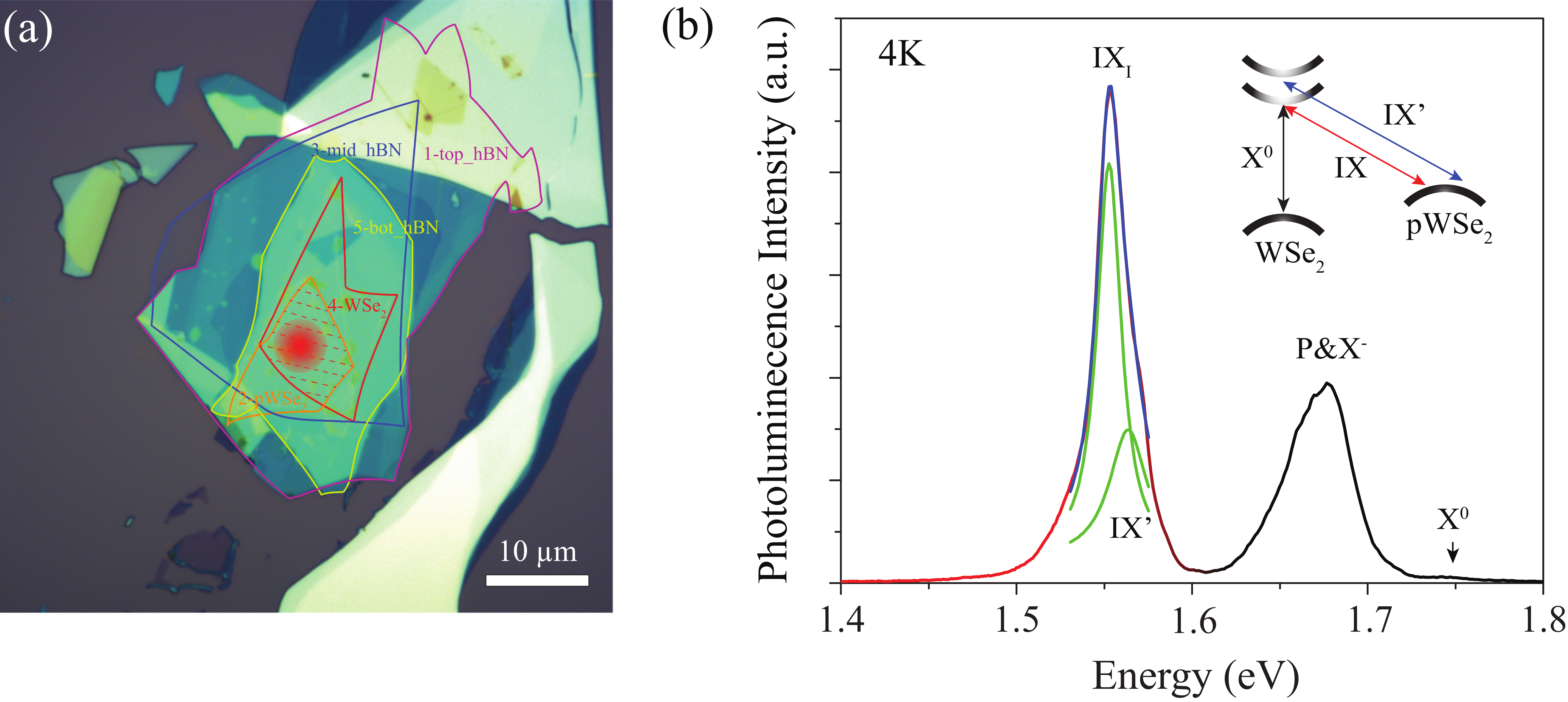}
	\caption{(a) Optical microscope image of interlayer excitons device, consisting of hBN/WSe$_2$/hBN/p:WSe$_2$/hBN. The flakes were transferred to a p-doped Si substrate with a top-down method. (b) A typical photoluminescence (PL) spectrum of the heterostructure sample taken at 4 K by He-Ne laser. This shows that at cryogenic temperatures, the PL from the interlayer excitons dominates the spectrum of the full heterostructure. A bi-Lorentzian fit is applied to the IX spectrum, which is consistent with an IX doublet with energy splitting $\Delta E = 20$ meV. Inset: illustration of the interlayer exciton transitions in the Type II structure, including the spin levels of the WSe$_2$ conduction band.}
	\label{fig.1}
\end{figure}

For this work we fabricated two quantum well-like stacked heterostrutures with the layer ordering hBN/WSe$_2$/hBN/p:WSe$_2$/hBN with different hBN thickness for the spacer layer between the two monolayers of WSe$_2$. We present direct observation of interlayer excitons in a type II band structure. We show that the IX energy blueshifts as the temperature increases. We find the populations of the doublet for IXs can be modified via tuning the external excitation laser power, a direct illustration of filling up the band energy level from the lower one. Moreover, a kink around 30 K of the integrated intensity ratio of the IXs to intralayer excitons divides the system into two regimes, namely thermal equilibrium above 30 K and non-equilibrium below 30 K. The lack of thermal equilibrium may indicate that dark (non-light-emitting) exciton states play a significant role when the temperature is below 30 K. Finally, although time-resolved PL measurements give an IX lifetime up to $\sim$ 790 ps, the decay rate can be considerable increased at higher density, which can be taken as evidence for an Auger recombination process. To the best of our knowledge, the Auger effect may not be avoidable in TMD heterostructures\cite{Wang2018a,Sun2014}, although inserting a thin layer of large bandgap semiconductor in between the monolayers of WSe$_2$ could weaken the Auger process without significantly reducing the interaction of the spatially separated electrons and holes.\\

\section{Results and discussion}
Isolated monolayer flakes of WSe$_2$ and p:WSe$_2$, with a typical size of 15 $\times$ 20 $\mu$m, were mechanically exfoliated from bulk crystals purchased from 2D Semiconductors. The two monolayers were encapsulated and separated by the hBN (grown at 4.5 GPa and 1500 C)\cite{TANIGUCHI2007}; the whole sandwich structure had the layer order  hBN/WSe$_2$/hBN/p:WSe$_2$/hBN from bottom to top. The samples were fabricated by standard dry transferring in a top-down method. We made two similar structure samples and one control sample consisting of hBN/WSe$_2$/hBN. All the samples had top and bottom hBN with a similar thickness of 15 nm, while the thickness of the hBN spacer layers were approximately 2 and 5 nm for samples 1 and 2 respectively. We focus on the data collected from sample 1 in this paper; the same measurements were performed on sample 2, which showed similar results as presented in the supplementary information. \\

Figure 1(a) gives an optical microscope image of the sample and Figure 1(b) gives a typical Photoluminescence (PL) spectrum of the sample. The inset illustrates that the p-doping of the intrinsic WSe$_2$ in the heterostructure raises the Fermi level above the valence band, which results in band renormalization that gives a type II structure. The PL was performed by non-resonantly pumping the device with a 632-nm He-Ne laser. The excitation light power was 0.07 mW with a spot size of 5 $\mu$m at normal incidence through a numerical aperture (NA) of 0.75 ($\times$50) microscope objective (Mitutoyo). The PL signal was collected by the same objective and directed to a grating spectrometer equipped with a charge-coupled device (CCD). All measurements were carried out by cooling the heterostructures to low temperature (below 10 K) in a continuous-flow cold-finger cryostat. Figure 1(b) shows a PL spectrum. As seen in this figure, the PL from the full heterostructure is dominated by the indirect exciton emission. Excitons trapped at impurities (labeled P) give a substantial contribution\cite{Wang2018a, Yan2014}. The indirect exciton emission at 1.568 eV is labeled IX, while emission from impurities and trions are both near 1.7 eV. We fit the IX peak with a sum of two Lorentzian lines to account for the spin-orbital splitting of the conduction band at the K point; the IX doublet has energy splitting $\Delta E=20$ meV \cite{Liu2013,Komider2013,Kormanyos2015}. The energy splitting agrees well with the theoretical calculation of the conduction band splitting for WSe$_2$, about 38 meV. In contrast to sample 2 with a spacer thickness of 5 nm, the IX energy here is about 82 meV lower, which is in agreement with a first-principles calculation \cite{Latini2017}. In general, the further the electron and the hole are from each other, the less the Coulomb interaction energy will be, and therefore the IX exciton binding energy will be less, which raises the absolute value of the IX energy. The results indicate that the interlayer coupling can be modified by the dielectric materials between the layers, providing an extra degree of control in the vdW heterostructure properties. Inset in Fig.1(b) illustrates the schematic of the lifted off conduction band of the WSe$_2$ and the double transitions channels in HS.\\

\begin{figure}[h]
	\centering
	\includegraphics[width=1\linewidth]{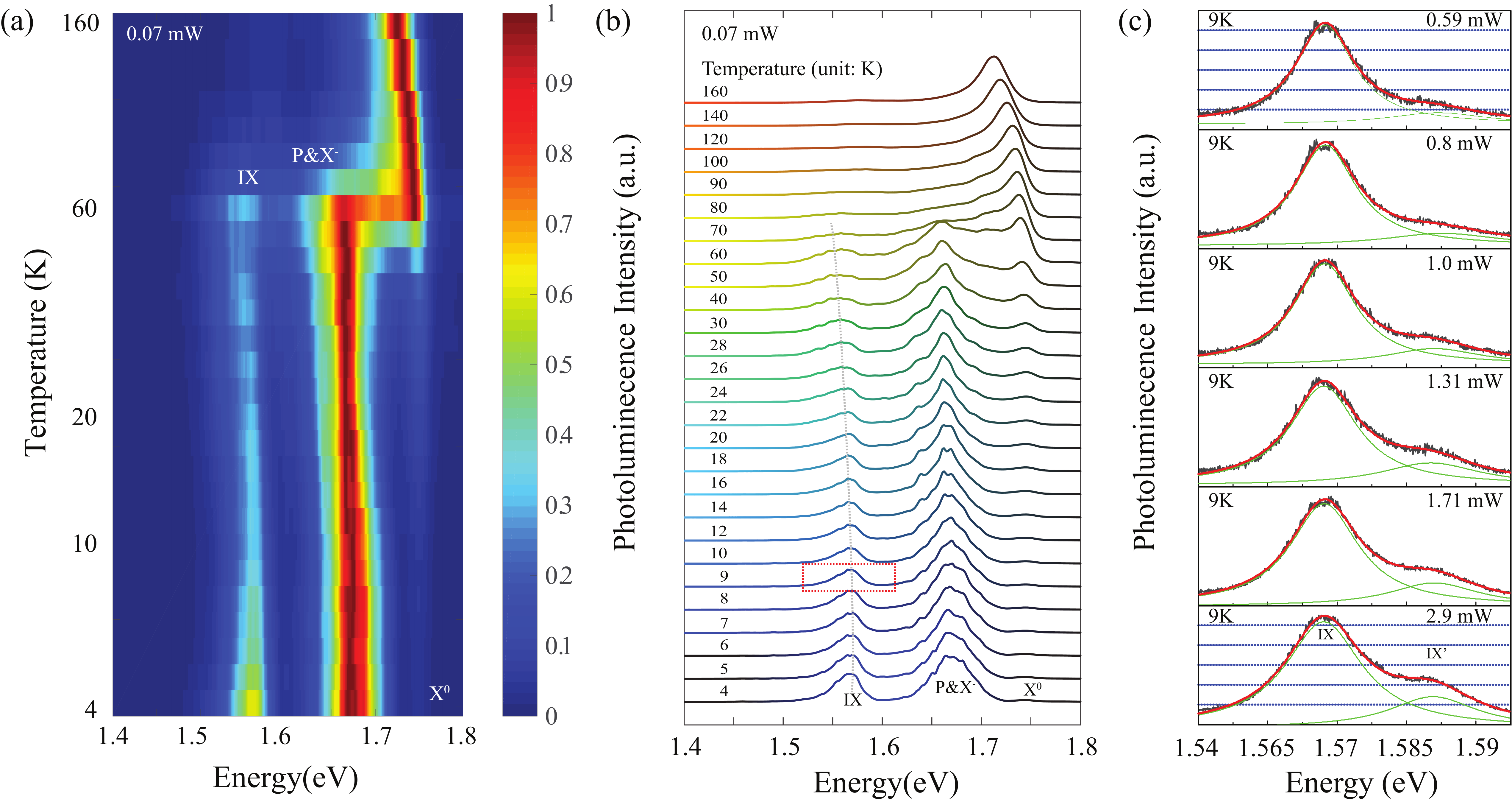}
	\caption{(a) (b) Two different visualizations of temperature-dependent PL created by a He-Ne pump laser with excitation power of 0.07 mW. The IX peak decreases as temperature increases and disappears around 60 K. The dashed gray line indicates the temperature-dependent bandgap shift. (c) Red curves: the power-dependent photoluminescence of interlayer excitons, for the spectral range shown in the box in (b), at 9 K for excitation by a mode-locked Ti:sapphire laser ($E_{\rm pump} \approx 1.75$ eV). Green curves: bi-Lorentzian fit to the plots, normalized for power and integration time. The blue grid lines help to demonstrate the relative intensities of the two peaks, and how they change with temperature. }
	\label{fig.2}
\end{figure}

\begin{figure}[!h]
	\centering
	\includegraphics[width=0.6\linewidth]{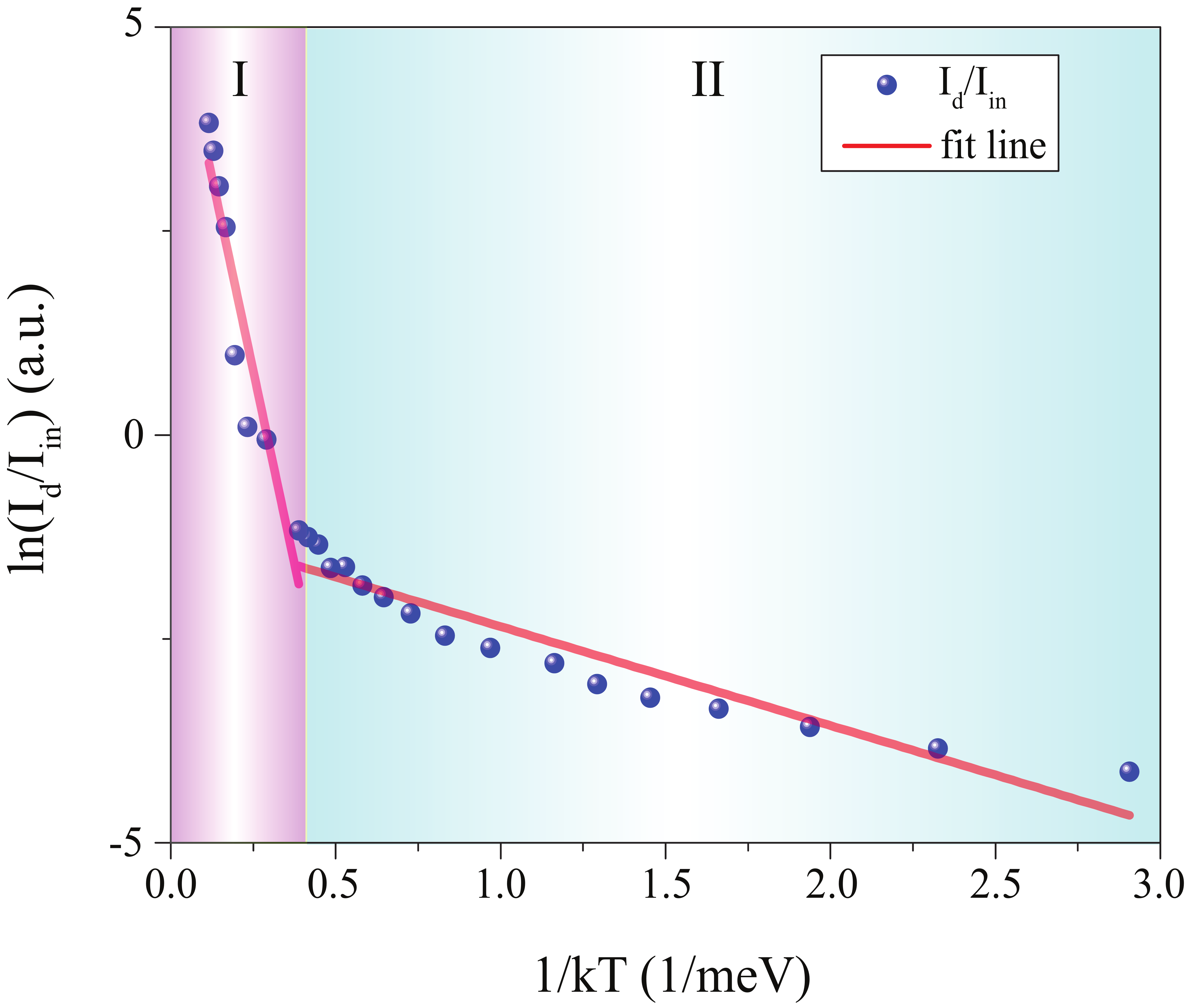}
	\caption{Natural log of the ratio of the integrated intensity of the direct and interlayer excitons as a function of 1/kT, where k is the Boltzmann constant. The colored lines give the results straight-line fits in the different temperature ranges, above and below 30 K. The fit slope in regime I and II are -190$\pm$17 meV and -12$\pm$1 meV, respectively.}
	\label{fig.3}
\end{figure}

\begin{figure}[!h]
	\centering
	\includegraphics[width=0.9\linewidth]{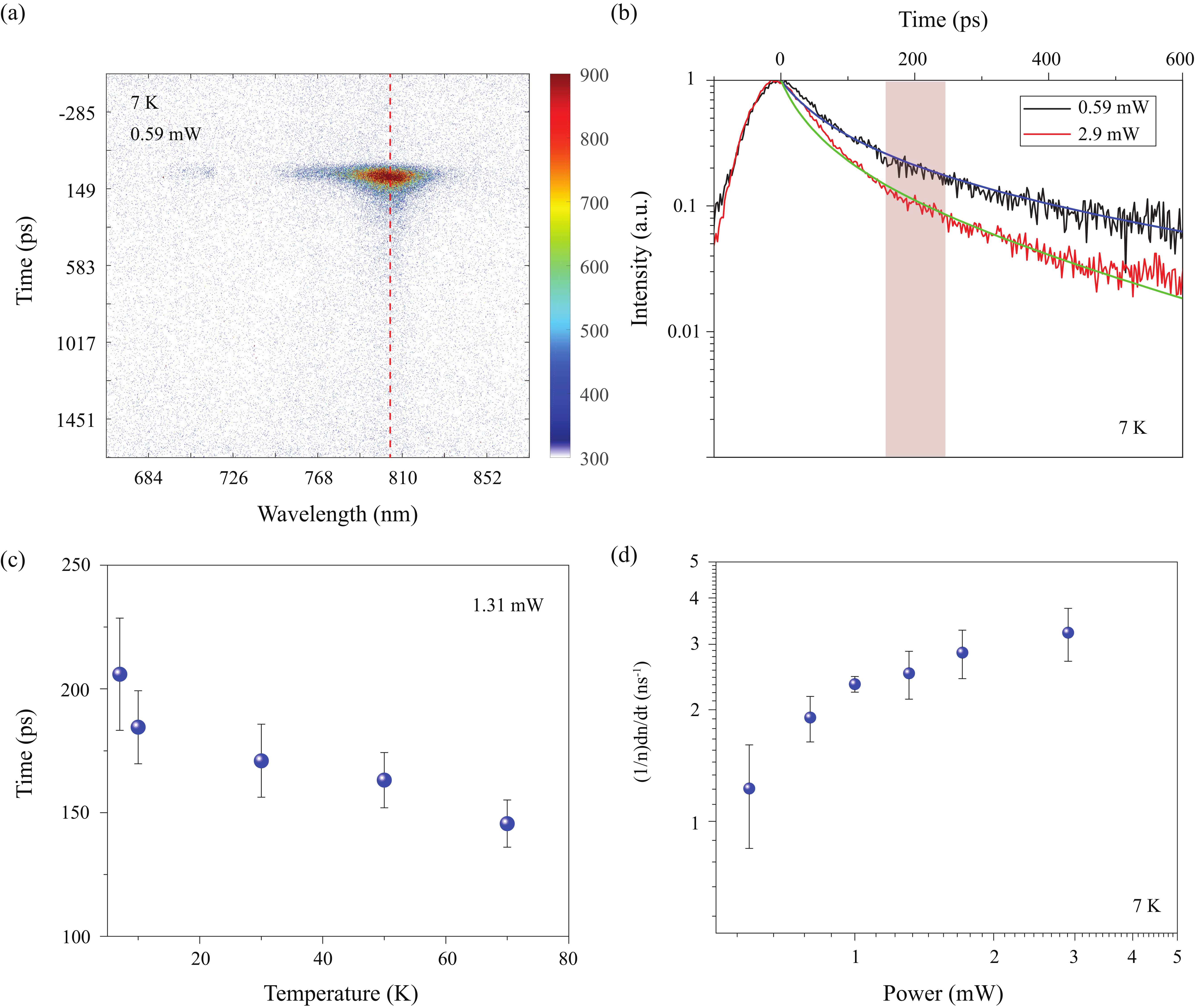}
	\caption{(a) Time-resolved photoluminescence image of HS (sample 1) measured by streak camera. The interlayer exciton (1.568 eV) shows a lifetime of $\sim$790 ps at late times under excitation power of 0.59 mW at 7 K. (b) Lifetime measurements for two different excitation power 0.59 mW and 2.9 mW at 7 K. $t=0$ is defined as the point of maximum PL. The time-dependent PL curves are fitted with equation (2). (c) Temperature-dependent lifetime of the IXs, with pump power 1.31 mW, for the time window shaded in brown in (b). (d) Decay rate as the function of the pump power, for the same time window, measured at 7 K.}
	\label{fig.4}
\end{figure}

Figure 2 shows the results of the temperature-dependent PL for the same structure. The pronounced IX peaks become attenuated as the temperature increases and quench gradually up to 60 K. The dashed gray line in Fig. 2(b) follows the shift of the band gap as a function of the temperature. Fig. 2(c) shows the IX emission at 9 K as the power of a mode-locked Ti: sapphire pump laser ($E_{\rm pump} \approx 1.75 $ eV) was varied. A bi-Lorentzian fit was applied to the data (black line), which was normalized by pump power and charge-coupled device (CCD) integration time. The higher energy and lower energy Lorentzians are shown with the two green lines. The relative intensity of these two sub-peaks changed as we increased the pump power. We define $I_{\rm\omega'}$ as the intensity of the higher-energy Lorentzian integrated over all energies, while $I_{\rm \omega}$ corresponds to the lower energy Lorentzian. The ratio of these two integrated quantities, $\eta =I_{\rm\omega'}/I_{\rm \omega}$, was found to be $7.2\%$ and $19.3\%$ at pump powers of 0.59 mW and 2.9 mW respectively. The power dependence of this ratio implies that the IXs predominantly occupy the lower energy state at low power but begin to occupy the upper state at high power, which may be due to the laser heating the sample and giving a higher effective temperature. A power-dependent blueshift of the IXs has been reported by two other groups \cite{Rivera2015,Wang2018a}; however, it is not seen here because the hBN spacer layer significantly decreased the IX\textquotesingle s density. \\

A further analysis was performed to measure the peak intensity of the intralayer and interlayer excitons as a function of the temperature. The result was fit with the thermal distribution function\\
\begin{equation}
\dfrac{I_{\rm d}}{I_{\rm in}}=e^{\rm{{-{\Delta E}/{k_BT}}}}.
\end{equation}
Here I$_{\rm d}$ and I$_{\rm in}$ are the peak intensity of direct (intralayer) and indirect (interlayer) excitons, respectively, and $k_B$ is the Boltzmann constant.  We find that the temperature dependence has two distinct regimes, separated by a kink around 30 K, as seen in the plot of the relative intensity as a function of $1/k_BT$ in Fig 3. Fitting the data with a line in regime I, which corresponds to temperatures above 30 K, gives $\Delta E = 190 \pm17 $ meV. This is consistent with the splitting between the two lines, measured directly at 100 K as 193 meV. The same fitting procedure in regime II (below 30 K) gives $\Delta E = 12 \pm 1$ meV, which is much lower than the measured energy splitting, around 182 meV at 9 K.  This may indicate that the exciton states are no longer in thermal equilibrium, e.g., because excitons in the upper state cannot relax efficiently into the lower state, and/or it may indicate that dark (non-light-emitting) exciton states play a significant role in thermalization below 30 K. This is plausible as dark excitons have been observed at cryogenic temperatures \cite{Zhou2017}, and the energy splitting of the dark and bright exciton in WSe$_2$ has been estimated as 16 meV\cite{Echeverry2016}. 

Time-resolved PL measurements were performed using a Hamamatsu streak camera with a temporal resolution of $\sim$1 ps, mounted on an exit port of an 0.5-m spectrometer equipped with a 50 grooves/mm reflective grating. A mode-locked pulsed Ti: sapphire laser set to wavelength 711 nm ($E_{\rm pump} \approx 1.75$ eV) was used for pumping the heterostructures. A 750 nm long-pass filter was placed in the signal's path to block the pump laser. Fig. 4(a) shows the wavelength-resolved streaked image, with calibrated time axis. The red dotted line is centered on the IX emission, around 790 nm. The time-resolved PL is plotted in Fig. 4(b) for T=7 K for two different pump powers, 0.59 mW and 2.9 mW. The time $t=0$ is defined as the point of maximum PL. The curvature of these plots at early times indicates that the decay process of the IXs is nontrivial.  We expect that the decay rate equation for the IX density is governed by two terms, 
\begin{equation}
\dfrac{1}{n}\dfrac{dn}{dt} = -An-\dfrac{1}{\tau},
\end{equation}
where $n$ is the IX density, $\tau$ is the lifetime and $A$ is an Auger coefficient, corresponding to a nonradiative process in which two excitons collide, and one recombines, giving its energy to ionizing the other exciton, which then reappears when the hot electron and hole lose enough energy to bind to each other again\cite{Snoke2014}. The second term describes radiative recombination, which dominates at low density. The radiative lifetime $\tau$ for the IXs at 7 K under excitation power of 0.59 mW was found as $\sim 790$ ps, by  linearly fitting the tail in Fig 4(b) at late times. The result is compared with fits of the whole time dependence using Equation (2) ($800\pm10$ ps).  The lifetime at a late time is not affected significantly by the Auger recombination, which means the Auger recombination dominates the initial decaying. Figure 4(c) shows the decay rate at intermediate times in the time range of 150 - 250 ps, after the initial thermalization of the carriers but while their density is still high, as a function of temperature, and Figure 4(d) shows the decay rate in the same time range, as a function of density. The increase of the decay rate with density is consistent with an Auger recombination mechanism, but may also be the result of the higher density leading to higher temperature. The decay process in sample 2 shows similar behavior (see supplementary information). {\tiny }

\section{Summary}

In this paper, we have shown that interlayer excitons can be observed in a vertical stacked structure of hBN/WSe$_2$/hBN/p:WSe$_2$/hBN, which is similar to a double quantum well. As expected, the energy position of the IX states is quite sensitive to the thickness of the spacer between the two layers. 

The photoluminescence of the IXs is fit by a bi-Lorentzian, showing two IX states with different energies.  The intensity ratio of the IXs and intralayer excitons follows the standard thermal distribution when the system is above 30 K. However, below 30 K there is a large discrepancy, leading us to believe that dark excitons play a significant role in low temperature thermalization. 

 The lifetime of the IXs is affected by the temperature and the excitation power. This indicates that not only direct excitons, but also indirect excitons, are subject to an Auger process, which the hBN spacer layer does not prevent. 

Much earlier work has studied indirect excitons in GaAs/AlGaAs double quantum well structures, including very long transport distance\cite{Snoke2005}. The current results indicate that many of these same results may be possible to reproduce in TMD-based heterostructures. The simplified sandwich structure of the multiple layers TMDs and the tunability of the IX energy either by modifying the thickness of the spacer or with the external electric field makes this structure appealling for embedding in an optical microcavity to create polariton states\cite{Liu2014c, Sun2017a}. It also may have implications for the search of IX condensation and new optoelectronics applications in 2D semiconductors.

\begin{acknowledgement}

This work was supported by the US Army Research Office under MURI award W911NF1710312 (Z.S., J.B, D.W.S.). K.W. and T.T. acknowledge support from the EMEXT Element Strategy Initiative to Form Core Research Center, Grant Number JPMXP0112101001 and the CREST(JPMJCR15F3), JST. B.M.H. and Q.C. were supported by the Department of Energy under the Early Career Award program (\#DE-SC0018115). \\
The authors declare no competing interests.

\end{acknowledgement}

\begin{suppinfo}

\begin{itemize}
\item The Supporting Information is available free of charge online.
The SI includes the schematic of the decay channels of the heterostructure; the Auger process analysis; temperature dependent spectra of the monolayer WSe$_2$, p:WSe$_2$, and a control sample with the same configuration as the one shows in the manuscript but possesses a different thickness of the spacer. 
\end{itemize}

\end{suppinfo}

\bibliography{reference}

\providecommand{\latin}[1]{#1}
\makeatletter
\providecommand{\doi}
  {\begingroup\let\do\@makeother\dospecials
  \catcode`\{=1 \catcode`\}=2 \doi@aux}
\providecommand{\doi@aux}[1]{\endgroup\texttt{#1}}
\makeatother
\providecommand*\mcitethebibliography{\thebibliography}
\csname @ifundefined\endcsname{endmcitethebibliography}
  {\let\endmcitethebibliography\endthebibliography}{}
\begin{mcitethebibliography}{34}
\providecommand*\natexlab[1]{#1}
\providecommand*\mciteSetBstSublistMode[1]{}
\providecommand*\mciteSetBstMaxWidthForm[2]{}
\providecommand*\mciteBstWouldAddEndPuncttrue
  {\def\EndOfBibitem{\unskip.}}
\providecommand*\mciteBstWouldAddEndPunctfalse
  {\let\EndOfBibitem\relax}
\providecommand*\mciteSetBstMidEndSepPunct[3]{}
\providecommand*\mciteSetBstSublistLabelBeginEnd[3]{}
\providecommand*\EndOfBibitem{}
\mciteSetBstSublistMode{f}
\mciteSetBstMaxWidthForm{subitem}{(\alph{mcitesubitemcount})}
\mciteSetBstSublistLabelBeginEnd
  {\mcitemaxwidthsubitemform\space}
  {\relax}
  {\relax}

\bibitem[Butov \latin{et~al.}(2002)Butov, Lai, Ivanov, Gossard, and
  Chemla]{Butov2002}
Butov,~L.~V.; Lai,~C.~W.; Ivanov,~A.~L.; Gossard,~A.~C.; Chemla,~D.~S. {Towards
  Bose-Einstein condensation of excitons in potential traps}. \emph{Nature}
  \textbf{2002}, \emph{417}, 47--52\relax
\mciteBstWouldAddEndPuncttrue
\mciteSetBstMidEndSepPunct{\mcitedefaultmidpunct}
{\mcitedefaultendpunct}{\mcitedefaultseppunct}\relax
\EndOfBibitem
\bibitem[V\"or\"os \latin{et~al.}(2006)V\"or\"os, Snoke, Pfeiffer, and
  West]{snoke2006}
V\"or\"os,~Z.; Snoke,~D.~W.; Pfeiffer,~L.; West,~K. Trapping Excitons in a
  Two-Dimensional In-Plane Harmonic Potential: Experimental Evidence for
  Equilibration of Indirect Excitons. \emph{Phys. Rev. Lett.} \textbf{2006},
  \emph{97}, 016803\relax
\mciteBstWouldAddEndPuncttrue
\mciteSetBstMidEndSepPunct{\mcitedefaultmidpunct}
{\mcitedefaultendpunct}{\mcitedefaultseppunct}\relax
\EndOfBibitem
\bibitem[Ross \latin{et~al.}(2014)Ross, Klement, Jones, Ghimire, Yan, Mandrus,
  Taniguchi, Watanabe, Kitamura, Yao, Cobden, and Xu]{Ross2014a}
Ross,~J.~S.; Klement,~P.; Jones,~A.~M.; Ghimire,~N.~J.; Yan,~J.;
  Mandrus,~D.~G.; Taniguchi,~T.; Watanabe,~K.; Kitamura,~K.; Yao,~W.;
  Cobden,~D.~H.; Xu,~X. {Electrically tunable excitonic light-emitting diodes
  based on monolayer WSe2 p-n junctions}. \emph{Nature Nanotechnology}
  \textbf{2014}, \emph{9}, 268--272\relax
\mciteBstWouldAddEndPuncttrue
\mciteSetBstMidEndSepPunct{\mcitedefaultmidpunct}
{\mcitedefaultendpunct}{\mcitedefaultseppunct}\relax
\EndOfBibitem
\bibitem[Koperski \latin{et~al.}(2015)Koperski, Nogajewski, Arora, Cherkez,
  Mallet, Veuillen, Marcus, Kossacki, and Potemski]{Koperski2015a}
Koperski,~M.; Nogajewski,~K.; Arora,~A.; Cherkez,~V.; Mallet,~P.;
  Veuillen,~J.~Y.; Marcus,~J.; Kossacki,~P.; Potemski,~M. {Single photon
  emitters in exfoliated WSe2 structures}. \emph{Nature Nanotechnology}
  \textbf{2015}, \emph{10}, 503--506\relax
\mciteBstWouldAddEndPuncttrue
\mciteSetBstMidEndSepPunct{\mcitedefaultmidpunct}
{\mcitedefaultendpunct}{\mcitedefaultseppunct}\relax
\EndOfBibitem
\bibitem[Schaibley \latin{et~al.}(2016)Schaibley, Yu, Clark, Rivera, Ross,
  Seyler, Yao, and Xu]{Schaibley2016}
Schaibley,~J.~R.; Yu,~H.; Clark,~G.; Rivera,~P.; Ross,~J.~S.; Seyler,~K.~L.;
  Yao,~W.; Xu,~X. {Valleytronics in 2D materials}. \emph{Nature Reviews
  Materials} \textbf{2016}, \emph{1}, 16055\relax
\mciteBstWouldAddEndPuncttrue
\mciteSetBstMidEndSepPunct{\mcitedefaultmidpunct}
{\mcitedefaultendpunct}{\mcitedefaultseppunct}\relax
\EndOfBibitem
\bibitem[Xu \latin{et~al.}(2014)Xu, Yao, Xiao, and Heinz]{Xu2014a}
Xu,~X.; Yao,~W.; Xiao,~D.; Heinz,~T.~F. {Spin and pseudospins in layered
  transition metal dichalcogenides}. \emph{Nature Physics} \textbf{2014},
  \emph{10}, 343--350\relax
\mciteBstWouldAddEndPuncttrue
\mciteSetBstMidEndSepPunct{\mcitedefaultmidpunct}
{\mcitedefaultendpunct}{\mcitedefaultseppunct}\relax
\EndOfBibitem
\bibitem[Lopez-Sanchez \latin{et~al.}(2013)Lopez-Sanchez, Lembke, Kayci,
  Radenovic, and Kis]{Lopez-Sanchez2013}
Lopez-Sanchez,~O.; Lembke,~D.; Kayci,~M.; Radenovic,~A.; Kis,~A.
  {Ultrasensitive photodetectors based on monolayer MoS 2}. \emph{Nature
  Nanotechnology} \textbf{2013}, \emph{8}, 497--501\relax
\mciteBstWouldAddEndPuncttrue
\mciteSetBstMidEndSepPunct{\mcitedefaultmidpunct}
{\mcitedefaultendpunct}{\mcitedefaultseppunct}\relax
\EndOfBibitem
\bibitem[Chernikov \latin{et~al.}(2014)Chernikov, Berkelbach, Hill, Rigosi, Li,
  Aslan, Reichman, Hybertsen, and Heinz]{Chernikov2014}
Chernikov,~A.; Berkelbach,~T.~C.; Hill,~H.~M.; Rigosi,~A.; Li,~Y.;
  Aslan,~O.~B.; Reichman,~D.~R.; Hybertsen,~M.~S.; Heinz,~T.~F. Exciton Binding
  Energy and Nonhydrogenic Rydberg Series in Monolayer ${\mathrm{WS}}_{2}$.
  \emph{Phys. Rev. Lett.} \textbf{2014}, \emph{113}, 076802\relax
\mciteBstWouldAddEndPuncttrue
\mciteSetBstMidEndSepPunct{\mcitedefaultmidpunct}
{\mcitedefaultendpunct}{\mcitedefaultseppunct}\relax
\EndOfBibitem
\bibitem[Ugeda \latin{et~al.}(2014)Ugeda, Bradley, Shi, {Da Jornada}, Zhang,
  Qiu, Ruan, Mo, Hussain, Shen, Wang, Louie, and Crommie]{Ugeda2014}
Ugeda,~M.~M.; Bradley,~A.~J.; Shi,~S.~F.; {Da Jornada},~F.~H.; Zhang,~Y.;
  Qiu,~D.~Y.; Ruan,~W.; Mo,~S.~K.; Hussain,~Z.; Shen,~Z.~X.; Wang,~F.;
  Louie,~S.~G.; Crommie,~M.~F. {Giant bandgap renormalization and excitonic
  effects in a monolayer transition metal dichalcogenide semiconductor}.
  \emph{Nature Materials} \textbf{2014}, \emph{13}, 1091--1095\relax
\mciteBstWouldAddEndPuncttrue
\mciteSetBstMidEndSepPunct{\mcitedefaultmidpunct}
{\mcitedefaultendpunct}{\mcitedefaultseppunct}\relax
\EndOfBibitem
\bibitem[Liu \latin{et~al.}(2014)Liu, Galfsky, Sun, Xia, Lin, Lee,
  K{\'{e}}na-Cohen, and Menon]{Liu2014c}
Liu,~X.; Galfsky,~T.; Sun,~Z.; Xia,~F.; Lin,~E.~C.; Lee,~Y.~H.;
  K{\'{e}}na-Cohen,~S.; Menon,~V.~M. {Strong light-matter coupling in
  two-dimensional atomic crystals}. \emph{Nature Photonics} \textbf{2014},
  \emph{9}, 30--34\relax
\mciteBstWouldAddEndPuncttrue
\mciteSetBstMidEndSepPunct{\mcitedefaultmidpunct}
{\mcitedefaultendpunct}{\mcitedefaultseppunct}\relax
\EndOfBibitem
\bibitem[Sun \latin{et~al.}(2017)Sun, Gu, Ghazaryan, Shotan, Considine, Dollar,
  Chakraborty, Liu, Ghaemi, K{\'{e}}na-Cohen, and Menon]{Sun2017a}
Sun,~Z.; Gu,~J.; Ghazaryan,~A.; Shotan,~Z.; Considine,~C.; Dollar,~M.;
  Chakraborty,~B.; Liu,~X.; Ghaemi,~P.; K{\'{e}}na-Cohen,~S.; Menon,~V.
  {Optical control of roomerature valley polaritons}. \emph{Nature Photonics}
  \textbf{2017}, \emph{11}, 491--496\relax
\mciteBstWouldAddEndPuncttrue
\mciteSetBstMidEndSepPunct{\mcitedefaultmidpunct}
{\mcitedefaultendpunct}{\mcitedefaultseppunct}\relax
\EndOfBibitem
\bibitem[Chakraborty \latin{et~al.}(2018)Chakraborty, Gu, Sun, Khatoniar,
  Bushati, Boehmke, Koots, and Menon]{Chakraborty2018a}
Chakraborty,~B.; Gu,~J.; Sun,~Z.; Khatoniar,~M.; Bushati,~R.; Boehmke,~A.~L.;
  Koots,~R.; Menon,~V.~M. {Control of Strong Light-Matter Interaction in
  Monolayer WS2 through Electric Field Gating}. \emph{Nano Letters}
  \textbf{2018}, \emph{18}, 6455--6460\relax
\mciteBstWouldAddEndPuncttrue
\mciteSetBstMidEndSepPunct{\mcitedefaultmidpunct}
{\mcitedefaultendpunct}{\mcitedefaultseppunct}\relax
\EndOfBibitem
\bibitem[Zhou \latin{et~al.}(2020)Zhou, Zhao, Tao, Li, Zhou, and
  Zhu]{Zhou2020a}
Zhou,~H.; Zhao,~Y.; Tao,~W.; Li,~Y.; Zhou,~Q.; Zhu,~H. {Controlling Exciton and
  Valley Dynamics in Two-Dimensional Heterostructures with Atomically Precise
  Interlayer Proximity}. \emph{ACS Nano} \textbf{2020}, \emph{14},
  4618--4625\relax
\mciteBstWouldAddEndPuncttrue
\mciteSetBstMidEndSepPunct{\mcitedefaultmidpunct}
{\mcitedefaultendpunct}{\mcitedefaultseppunct}\relax
\EndOfBibitem
\bibitem[Rivera \latin{et~al.}(2015)Rivera, Schaibley, Jones, Ross, Wu,
  Aivazian, Klement, Seyler, Clark, Ghimire, Yan, Mandrus, Yao, and
  Xu]{Rivera2015}
Rivera,~P.; Schaibley,~J.~R.; Jones,~A.~M.; Ross,~J.~S.; Wu,~S.; Aivazian,~G.;
  Klement,~P.; Seyler,~K.; Clark,~G.; Ghimire,~N.~J.; Yan,~J.; Mandrus,~D.~G.;
  Yao,~W.; Xu,~X. {Observation of long-lived interlayer excitons in monolayer
  MoSe 2-WSe 2 heterostructures}. \emph{Nature Communications} \textbf{2015},
  \emph{6}, 4--9\relax
\mciteBstWouldAddEndPuncttrue
\mciteSetBstMidEndSepPunct{\mcitedefaultmidpunct}
{\mcitedefaultendpunct}{\mcitedefaultseppunct}\relax
\EndOfBibitem
\bibitem[Hong \latin{et~al.}(2014)Hong, Kim, Shi, Zhang, Jin, Sun, Tongay, Wu,
  Zhang, and Wang]{Hong2014}
Hong,~X.; Kim,~J.; Shi,~S.~F.; Zhang,~Y.; Jin,~C.; Sun,~Y.; Tongay,~S.; Wu,~J.;
  Zhang,~Y.; Wang,~F. {Ultrafast charge transfer in atomically thin MoS2/WS2
  heterostructures}. \emph{Nature Nanotechnology} \textbf{2014}, \emph{9},
  682--686\relax
\mciteBstWouldAddEndPuncttrue
\mciteSetBstMidEndSepPunct{\mcitedefaultmidpunct}
{\mcitedefaultendpunct}{\mcitedefaultseppunct}\relax
\EndOfBibitem
\bibitem[Rivera \latin{et~al.}(2016)Rivera, Seyler, Yu, Schaibley, Yan,
  Mandrus, Yao, and Xu]{Rivera2016a}
Rivera,~P.; Seyler,~K.~L.; Yu,~H.; Schaibley,~J.~R.; Yan,~J.; Mandrus,~D.~G.;
  Yao,~W.; Xu,~X. {Valley-polarized exciton dynamics in a 2D semiconductor
  heterostructure}. \emph{Science} \textbf{2016}, \emph{351}, 688--691\relax
\mciteBstWouldAddEndPuncttrue
\mciteSetBstMidEndSepPunct{\mcitedefaultmidpunct}
{\mcitedefaultendpunct}{\mcitedefaultseppunct}\relax
\EndOfBibitem
\bibitem[Lee \latin{et~al.}(2014)Lee, Lee, {Van Der Zande}, Chen, Li, Han, Cui,
  Arefe, Nuckolls, Heinz, Guo, Hone, and Kim]{Lee2014}
Lee,~C.~H.; Lee,~G.~H.; {Van Der Zande},~A.~M.; Chen,~W.; Li,~Y.; Han,~M.;
  Cui,~X.; Arefe,~G.; Nuckolls,~C.; Heinz,~T.~F.; Guo,~J.; Hone,~J.; Kim,~P.
  {Atomically thin p-n junctions with van der Waals heterointerfaces}.
  \emph{Nature Nanotechnology} \textbf{2014}, \emph{9}, 676--681\relax
\mciteBstWouldAddEndPuncttrue
\mciteSetBstMidEndSepPunct{\mcitedefaultmidpunct}
{\mcitedefaultendpunct}{\mcitedefaultseppunct}\relax
\EndOfBibitem
\bibitem[Karni \latin{et~al.}(2019)Karni, Barr\'e, Lau, Gillen, Ma, Kim,
  Watanabe, Taniguchi, Maultzsch, Barmak, Page, and Heinz]{Karni2019}
Karni,~O.; Barr\'e,~E.; Lau,~S.~C.; Gillen,~R.; Ma,~E.~Y.; Kim,~B.;
  Watanabe,~K.; Taniguchi,~T.; Maultzsch,~J.; Barmak,~K.; Page,~R.~H.;
  Heinz,~T.~F. Infrared Interlayer Exciton Emission in
  ${\mathrm{MoS}}_{2}/{\mathrm{WSe}}_{2}$ Heterostructures. \emph{Phys. Rev.
  Lett.} \textbf{2019}, \emph{123}, 247402\relax
\mciteBstWouldAddEndPuncttrue
\mciteSetBstMidEndSepPunct{\mcitedefaultmidpunct}
{\mcitedefaultendpunct}{\mcitedefaultseppunct}\relax
\EndOfBibitem
\bibitem[Unuchek \latin{et~al.}(2019)Unuchek, Ciarrocchi, Avsar, Sun, Watanabe,
  Taniguchi, and Kis]{Unuchek2019}
Unuchek,~D.; Ciarrocchi,~A.; Avsar,~A.; Sun,~Z.; Watanabe,~K.; Taniguchi,~T.;
  Kis,~A. {Valley-polarized exciton currents in a van der Waals
  heterostructure}. \emph{Nature Nanotechnology} \textbf{2019}, \emph{14},
  1104--1109\relax
\mciteBstWouldAddEndPuncttrue
\mciteSetBstMidEndSepPunct{\mcitedefaultmidpunct}
{\mcitedefaultendpunct}{\mcitedefaultseppunct}\relax
\EndOfBibitem
\bibitem[Wang \latin{et~al.}(2018)Wang, Chiu, Honz, Mak, and Shan]{Wang2018a}
Wang,~Z.; Chiu,~Y.~H.; Honz,~K.; Mak,~K.~F.; Shan,~J. {Electrical Tuning of
  Interlayer Exciton Gases in WSe$_2$ Bilayers}. \emph{Nano Letters}
  \textbf{2018}, \emph{18}, 137--143\relax
\mciteBstWouldAddEndPuncttrue
\mciteSetBstMidEndSepPunct{\mcitedefaultmidpunct}
{\mcitedefaultendpunct}{\mcitedefaultseppunct}\relax
\EndOfBibitem
\bibitem[Fang \latin{et~al.}(2014)Fang, Battaglia, Carraro, Nemsak, Ozdol,
  Kang, Bechtel, Desai, Kronast, Unal, Conti, Conlon, Palsson, Martin, Minor,
  Fadley, Yablonovitch, Maboudian, and Javey]{Fang2014}
Fang,~H. \latin{et~al.}  Strong interlayer coupling in van der Waals
  heterostructures built from single-layer chalcogenides. \emph{Proceedings of
  the National Academy of Sciences} \textbf{2014}, \emph{111}, 6198--6202\relax
\mciteBstWouldAddEndPuncttrue
\mciteSetBstMidEndSepPunct{\mcitedefaultmidpunct}
{\mcitedefaultendpunct}{\mcitedefaultseppunct}\relax
\EndOfBibitem
\bibitem[Calman \latin{et~al.}(2018)Calman, Fogler, Butov, Hu, Mishchenko, and
  Geim]{Calman2018}
Calman,~E.~V.; Fogler,~M.~M.; Butov,~L.~V.; Hu,~S.; Mishchenko,~A.; Geim,~A.~K.
  {Indirect excitons in van der Waals heterostructures at room temperature}.
  \emph{Nature Communications} \textbf{2018}, \emph{9}, 1--5\relax
\mciteBstWouldAddEndPuncttrue
\mciteSetBstMidEndSepPunct{\mcitedefaultmidpunct}
{\mcitedefaultendpunct}{\mcitedefaultseppunct}\relax
\EndOfBibitem
\bibitem[Latini \latin{et~al.}(2017)Latini, Winther, Olsen, and
  Thygesen]{Latini2017}
Latini,~S.; Winther,~K.~T.; Olsen,~T.; Thygesen,~K.~S. {Interlayer Excitons and
  Band Alignment in MoS$_2$/hBN/WSe$_2$ van der Waals Heterostructures}.
  \emph{Nano Letters} \textbf{2017}, \emph{17}, 938--945\relax
\mciteBstWouldAddEndPuncttrue
\mciteSetBstMidEndSepPunct{\mcitedefaultmidpunct}
{\mcitedefaultendpunct}{\mcitedefaultseppunct}\relax
\EndOfBibitem
\bibitem[Sun \latin{et~al.}(2014)Sun, Rao, Reider, Chen, You, Br{\'{e}}zin,
  Harutyunyan, and Heinz]{Sun2014}
Sun,~D.; Rao,~Y.; Reider,~G.~A.; Chen,~G.; You,~Y.; Br{\'{e}}zin,~L.;
  Harutyunyan,~A.~R.; Heinz,~T.~F. {Observation of rapid exciton-exciton
  annihilation in monolayer molybdenum disulfide}. \emph{Nano Letters}
  \textbf{2014}, \emph{14}, 5625--5629\relax
\mciteBstWouldAddEndPuncttrue
\mciteSetBstMidEndSepPunct{\mcitedefaultmidpunct}
{\mcitedefaultendpunct}{\mcitedefaultseppunct}\relax
\EndOfBibitem
\bibitem[Taniguchi and Watanabe(2007)Taniguchi, and Watanabe]{TANIGUCHI2007}
Taniguchi,~T.; Watanabe,~K. Synthesis of high-purity boron nitride single
  crystals under high pressure by using Ba–BN solvent. \emph{Journal of
  Crystal Growth} \textbf{2007}, \emph{303}, 525 -- 529\relax
\mciteBstWouldAddEndPuncttrue
\mciteSetBstMidEndSepPunct{\mcitedefaultmidpunct}
{\mcitedefaultendpunct}{\mcitedefaultseppunct}\relax
\EndOfBibitem
\bibitem[Yan \latin{et~al.}(2014)Yan, Qiao, Liu, Tan, and Zhang]{Yan2014}
Yan,~T.; Qiao,~X.; Liu,~X.; Tan,~P.; Zhang,~X. { Photoluminescence properties
  and exciton dynamics in monolayer WSe 2 }. \emph{Applied Physics Letters}
  \textbf{2014}, \emph{105}, 101901\relax
\mciteBstWouldAddEndPuncttrue
\mciteSetBstMidEndSepPunct{\mcitedefaultmidpunct}
{\mcitedefaultendpunct}{\mcitedefaultseppunct}\relax
\EndOfBibitem
\bibitem[Liu \latin{et~al.}(2013)Liu, Shan, Yao, Yao, and Xiao]{Liu2013}
Liu,~G.~B.; Shan,~W.~Y.; Yao,~Y.; Yao,~W.; Xiao,~D. {Three-band tight-binding
  model for monolayers of group-VIB transition metal dichalcogenides}.
  \emph{Physical Review B} \textbf{2013}, \emph{88}, 085433\relax
\mciteBstWouldAddEndPuncttrue
\mciteSetBstMidEndSepPunct{\mcitedefaultmidpunct}
{\mcitedefaultendpunct}{\mcitedefaultseppunct}\relax
\EndOfBibitem
\bibitem[Komider \latin{et~al.}(2013)Komider, Gonz{\'{a}}lez, and
  Fern{\'{a}}ndez-Rossier]{Komider2013}
Komider,~K.; Gonz{\'{a}}lez,~J.~W.; Fern{\'{a}}ndez-Rossier,~J. {Large spin
  splitting in the conduction band of transition metal dichalcogenide
  monolayers}. \emph{Physical Review B} \textbf{2013}, \emph{88}, 245436\relax
\mciteBstWouldAddEndPuncttrue
\mciteSetBstMidEndSepPunct{\mcitedefaultmidpunct}
{\mcitedefaultendpunct}{\mcitedefaultseppunct}\relax
\EndOfBibitem
\bibitem[Korm{\'{a}}nyos \latin{et~al.}(2015)Korm{\'{a}}nyos, Burkard, Gmitra,
  Fabian, Z{\'{o}}lyomi, Drummond, and Fal]{Kormanyos2015}
Korm{\'{a}}nyos,~A.; Burkard,~G.; Gmitra,~M.; Fabian,~J.; Z{\'{o}}lyomi,~V.;
  Drummond,~N.~D.; Fal,~V. {Erratum: K{\textperiodcentered}p theory for
  two-dimensional transition metal dichalcogenide semiconductors}. \emph{2D
  Materials} \textbf{2015}, \emph{2}, 22001\relax
\mciteBstWouldAddEndPuncttrue
\mciteSetBstMidEndSepPunct{\mcitedefaultmidpunct}
{\mcitedefaultendpunct}{\mcitedefaultseppunct}\relax
\EndOfBibitem
\bibitem[Zhou \latin{et~al.}(2017)Zhou, Scuri, Wild, High, Dibos, Jauregui,
  Shu, {De Greve}, Pistunova, Joe, Taniguchi, Watanabe, Kim, Lukin, and
  Park]{Zhou2017}
Zhou,~Y.; Scuri,~G.; Wild,~D.~S.; High,~A.~A.; Dibos,~A.; Jauregui,~L.~A.;
  Shu,~C.; {De Greve},~K.; Pistunova,~K.; Joe,~A.~Y.; Taniguchi,~T.;
  Watanabe,~K.; Kim,~P.; Lukin,~M.~D.; Park,~H. {Probing dark excitons in
  atomically thin semiconductors via near-field coupling to surface plasmon
  polaritons}. \emph{Nature Nanotechnology} \textbf{2017}, \emph{12},
  856--860\relax
\mciteBstWouldAddEndPuncttrue
\mciteSetBstMidEndSepPunct{\mcitedefaultmidpunct}
{\mcitedefaultendpunct}{\mcitedefaultseppunct}\relax
\EndOfBibitem
\bibitem[Echeverry \latin{et~al.}(2016)Echeverry, Urbaszek, Amand, Marie, and
  Gerber]{Echeverry2016}
Echeverry,~J.~P.; Urbaszek,~B.; Amand,~T.; Marie,~X.; Gerber,~I.~C. Splitting
  between bright and dark excitons in transition metal dichalcogenide
  monolayers. \emph{Phys. Rev. B} \textbf{2016}, \emph{93}, 121107\relax
\mciteBstWouldAddEndPuncttrue
\mciteSetBstMidEndSepPunct{\mcitedefaultmidpunct}
{\mcitedefaultendpunct}{\mcitedefaultseppunct}\relax
\EndOfBibitem
\bibitem[Snoke and Kavoulakis(2014)Snoke, and Kavoulakis]{Snoke2014}
Snoke,~D.; Kavoulakis,~G.~M. Bose{\textendash}Einstein condensation of excitons
  in Cu2O: progress over 30 years. \emph{Reports on Progress in Physics}
  \textbf{2014}, \emph{77}, 116501\relax
\mciteBstWouldAddEndPuncttrue
\mciteSetBstMidEndSepPunct{\mcitedefaultmidpunct}
{\mcitedefaultendpunct}{\mcitedefaultseppunct}\relax
\EndOfBibitem
\bibitem[V\"or\"os \latin{et~al.}(2005)V\"or\"os, Balili, Snoke, Pfeiffer, and
  West]{Snoke2005}
V\"or\"os,~Z.; Balili,~R.; Snoke,~D.~W.; Pfeiffer,~L.; West,~K. Long-Distance
  Diffusion of Excitons in Double Quantum Well Structures. \emph{Phys. Rev.
  Lett.} \textbf{2005}, \emph{94}, 226401\relax
\mciteBstWouldAddEndPuncttrue
\mciteSetBstMidEndSepPunct{\mcitedefaultmidpunct}
{\mcitedefaultendpunct}{\mcitedefaultseppunct}\relax
\EndOfBibitem
\end{mcitethebibliography}

\newpage
\centering
\textbf{TOC Graphic}
\begin{figure}[!h]
	\centering
	\includegraphics[width=1\linewidth]{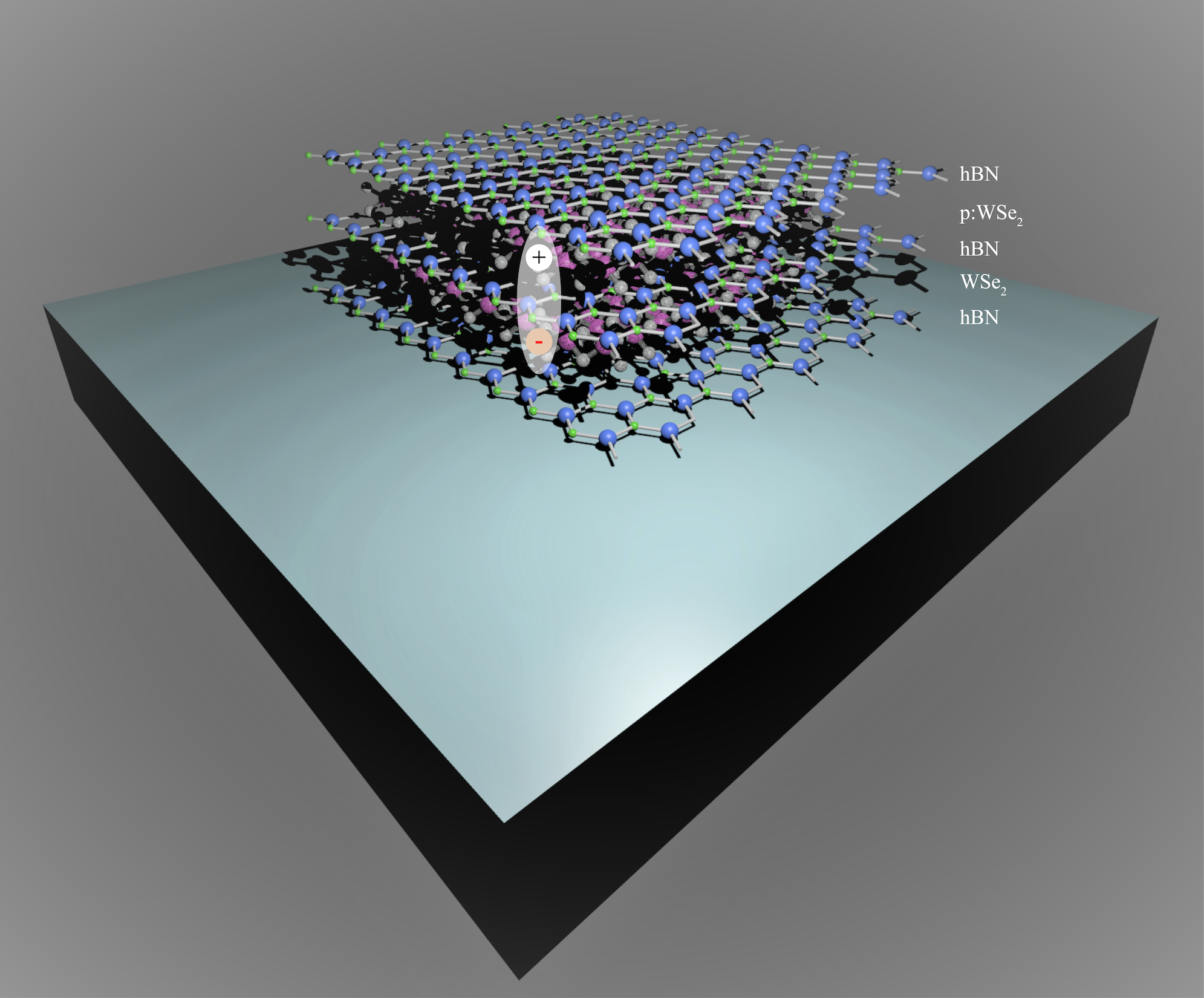}
\end{figure}

\end{document}


\newpage
\begin{center}
	I Thermal Distribution
\end{center}
\begin{figure}[H]
	\centering
	\includegraphics[width=0.6\linewidth]{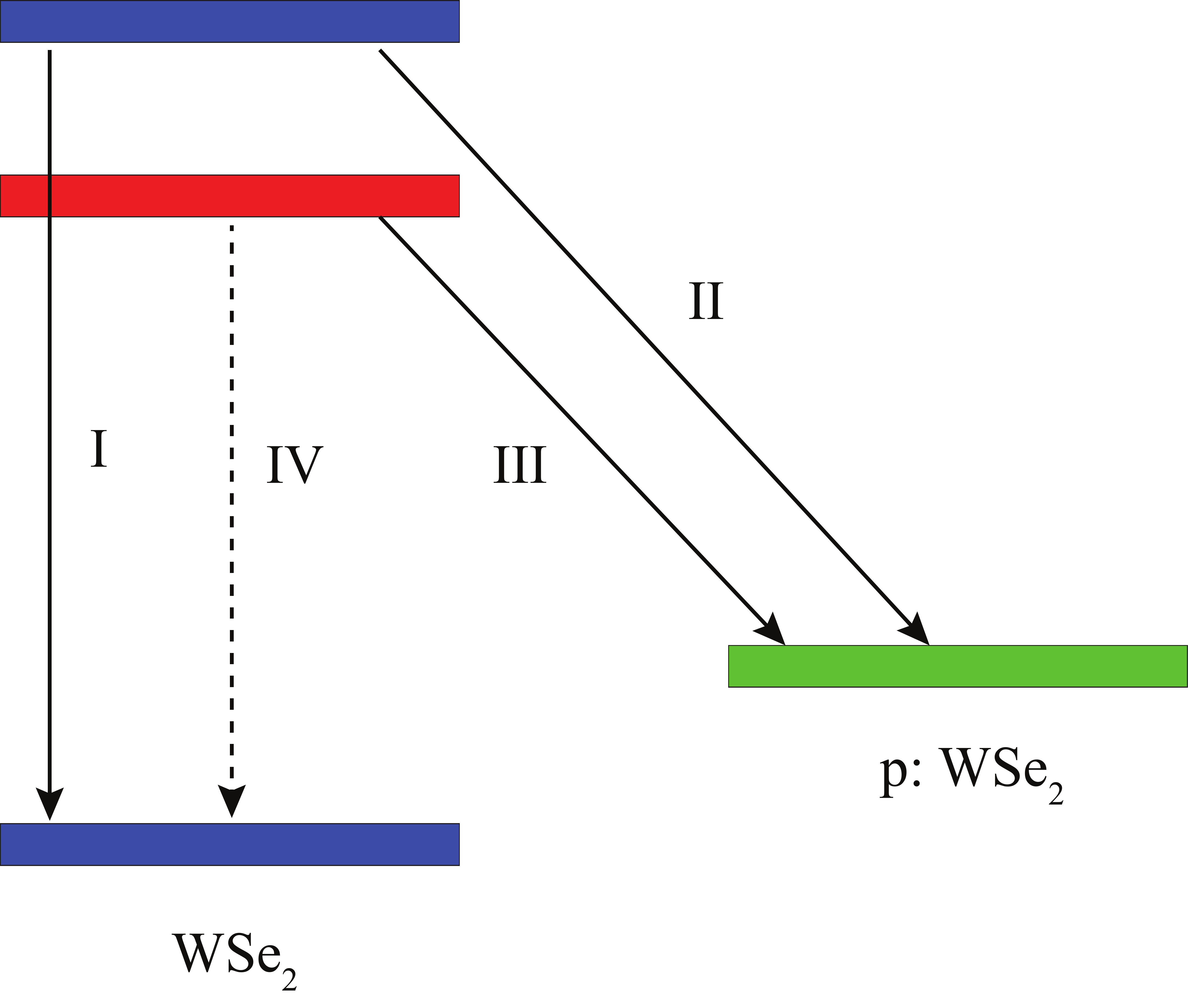}
	\caption{Schemetic of the decay channels}
	\label{S:SI1}
\end{figure}
The carriers are following the thermal distribution in the sample. The total number of the carriers is N$_0$.
 At the temperature above the 30K, the I channel dominates the whole spectrum. II and III are weaker, IV channel is almost forbidden due to its ultra-long lifetime. Thus, the intensity ratio of the intralayer and interlayer exciton can be described as:
\\
\begin{equation}
\dfrac{I_d}{I_{in}}=\dfrac{N_0exp(-E_I/k_BT)}{N_0exp(-E_{II or III}/k_BT)}=exp(-(E_I-E_{II or III})/k_BT)
\end{equation}
When the temperature goes below 30K, the channel IV is reactivated, due to more carriers are inclinedly to fall into the lower level (dark state) because of the phonon can not efficiently excite them up to the higher level (bright state).  Thus, the ratio of the intensity can be written as:
\begin{equation}
\dfrac{I_d}{I_{in}}=\dfrac{N_0exp(-E_I/k_BT)(1-\gamma(T))}{N_0exp(-E_{II or III}/k_BT)}
\end{equation}
$\gamma(T)$ is a constant with no significant temperature-dependent in our experiment and is about 0.95 for both samples.

\begin{center}
	II Auger Process
\end{center}
We describe the lifetime of the interlayer exciton by using the decay rate equation below:
\begin{equation}
\dfrac{1}{n}\dfrac{dn}{dt} = -An-\dfrac{1}{\tau},
\end{equation}
The first term on the right is the Auger recombination process with the coefficient A. The second term designates the exciton recombination with the lifetime of $\tau$. 
We fit the data of the figure 4 with the decay differential equation. And the fitting data are list here:
\begin{table}
	\caption{Auger process fitting}
	\label{Auger process fitting}
	\begin{tabular}{|c|c|c|}
		\hline
		Pump power (mW)  & Auger coefficient(A)(/ps) & Lifetime($\tau$)(ps) \\
		\hline
		0.59   & 0.015$\pm$0.0015 &  800$\pm$10\\
		\hline
	    2.9 & 0.025$\pm$0.002  & 320$\pm$12\\
        \hline
	\end{tabular}
\end{table}
The Auger coefficient increases with the pump power are straightforward to understand. As we expected that the lifetime does not show significant sensitivity to the first right term and thus can be fit by the tail of the data as we discussed in the manuscript. 

\begin{figure}[H]
	\centering
	\includegraphics[width=0.6\linewidth]{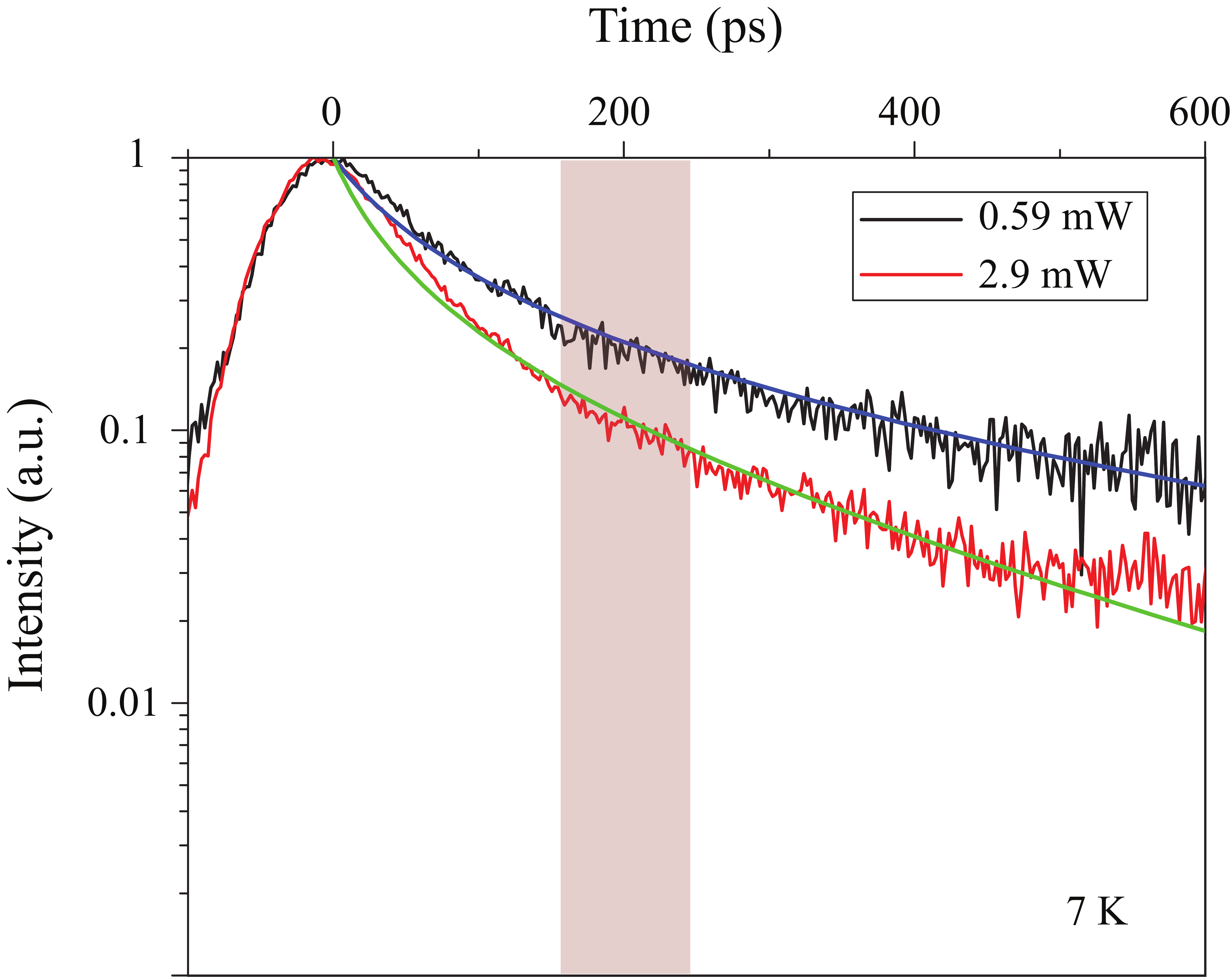}
	\caption{Lifetime measurements for two different excitation power 0.59 mW and 2.9 mW at 7 K. t = 0 is defined as the point of maximum PL. The
whole spectra are fitted with equation 3.}
	\label{S:SI2}
\end{figure}

\newpage
\begin{center}
	III. Control Sample 1: monolayer hBN/WSe$_2$/hBN
\end{center}

\begin{figure}[H]
	\centering
	\includegraphics[width=0.9\linewidth]{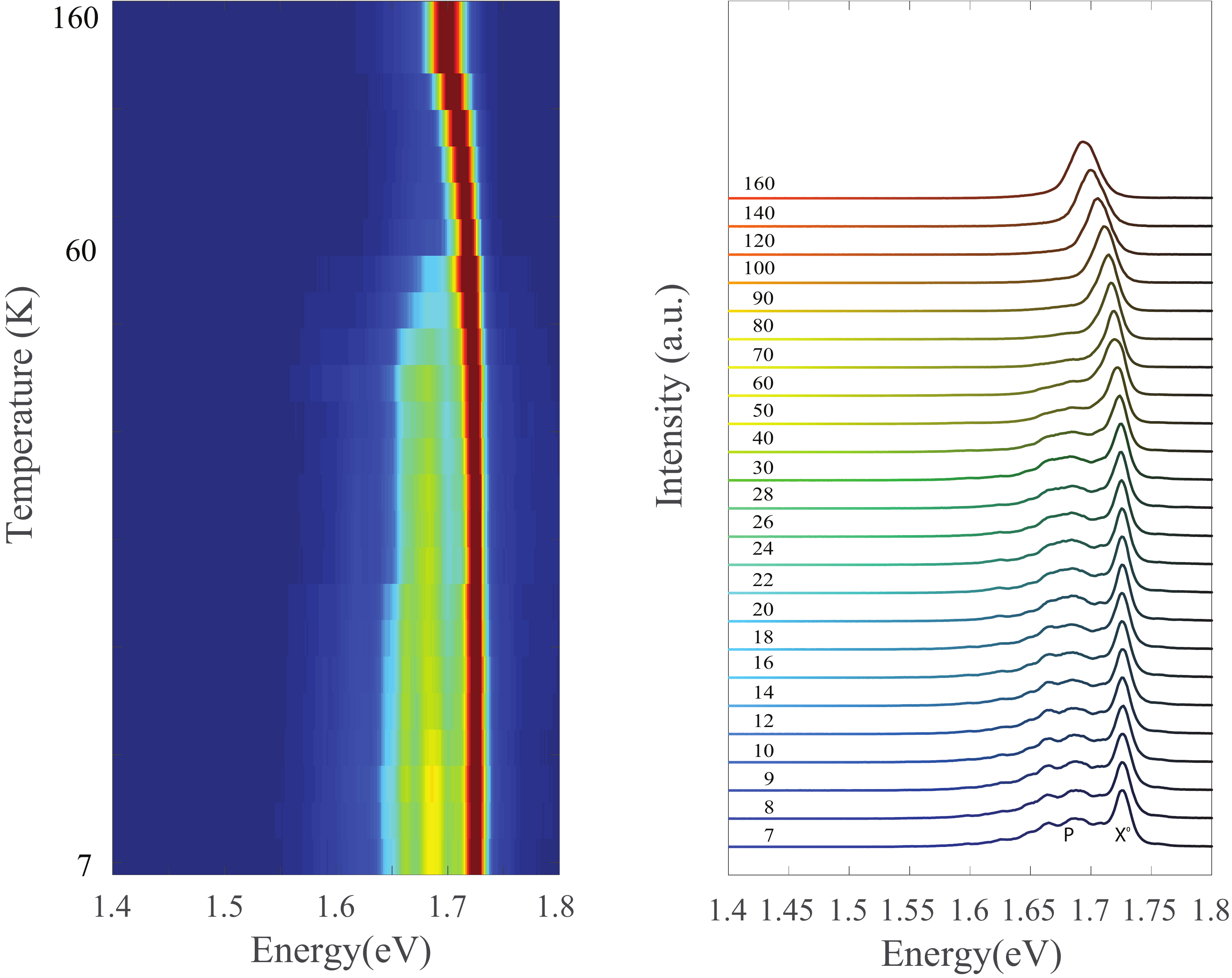}
	\caption{Temperature dependent photoluminescence of the monolayer WSe$_2$ encapsulated with hBN. }
	\label{S:S3}
\end{figure}

Figure S3 shows the temperature-dependent photolumnscence spectrum monolayer WSe$_2$ encapsulated in hBN. The line labeled X$^0$ corresponds to the intrinsic excitons while the line labeled P is the emission of excitons bound to impurities; trion emission also occurs in this range. \\

\newpage
\begin{center}
	IV. Control Sample 2: hBN/p:WSe$_2$/hBN
\end{center}

\begin{figure}[H]
	\centering
	\includegraphics[width=0.9\linewidth]{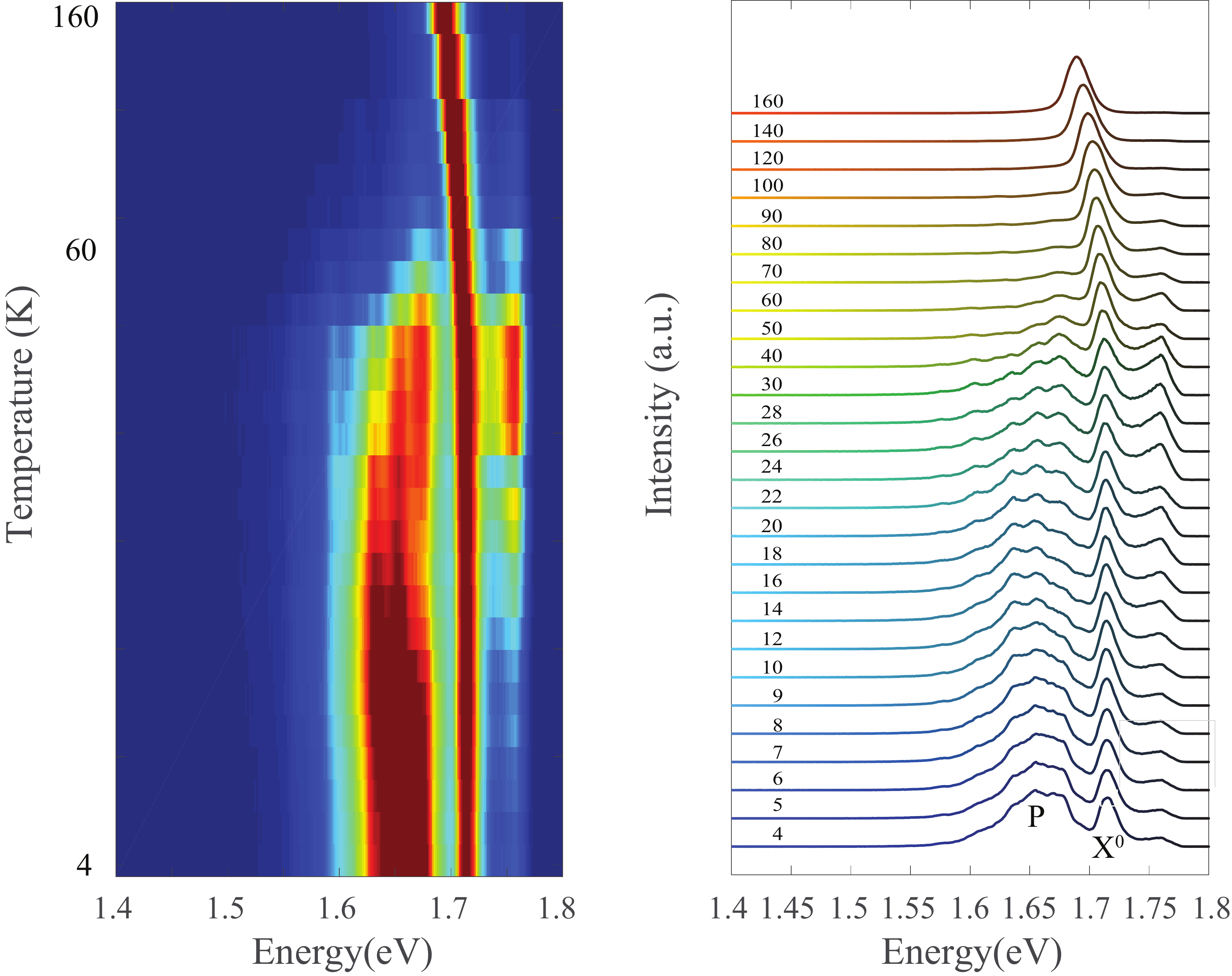}
	\caption{Temperature dependent photoluminescence of the monolayer p:WSe$_2$ encapsulated with hBN. }
	\label{S:SI4}
\end{figure}

Figure S4 shows the same type of data as Figure S3, but with a doped sample, namely p:WSe$_2$ encapsulated in hBN.  The peak above 1.75 eV is an artifact of scattered laser light cut off by a 700-nm long-pass filter.
\\

\newpage
\begin{center}
	V. Sample 2: hBN/WSe$_2$/hBN/p:WSe$_2$/hBN
\end{center}
\begin{figure}[H]
	\centering
	\includegraphics[width=1\linewidth]{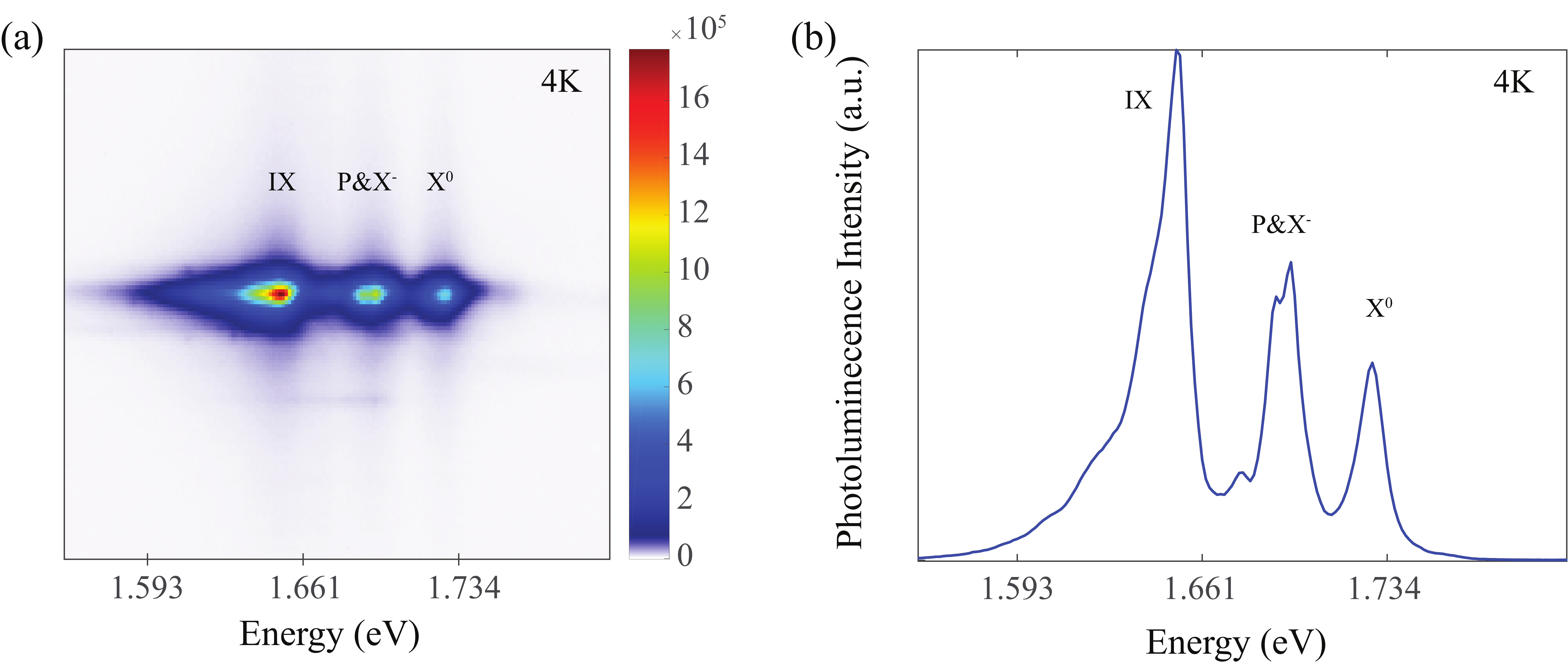}
	\caption{A typical photoluminescence (PL) spectrum (spatially resolved, in (a), and spatially integrated, in (b)) of the heterostructure of Sample 2 (bilayer) taken at 4 K by He-Ne pump laser with an excitation power of 0.07 mW.  All three peaks are pronounced at cryogenic temperature. Here the energy of the interlayer excitons and intralayer energies are 1.65 eV and 1.73 eV, respectively.}
	\label{S:SI5}
\end{figure}

\begin{figure}[H]
	\centering
	\includegraphics[width=0.9\linewidth]{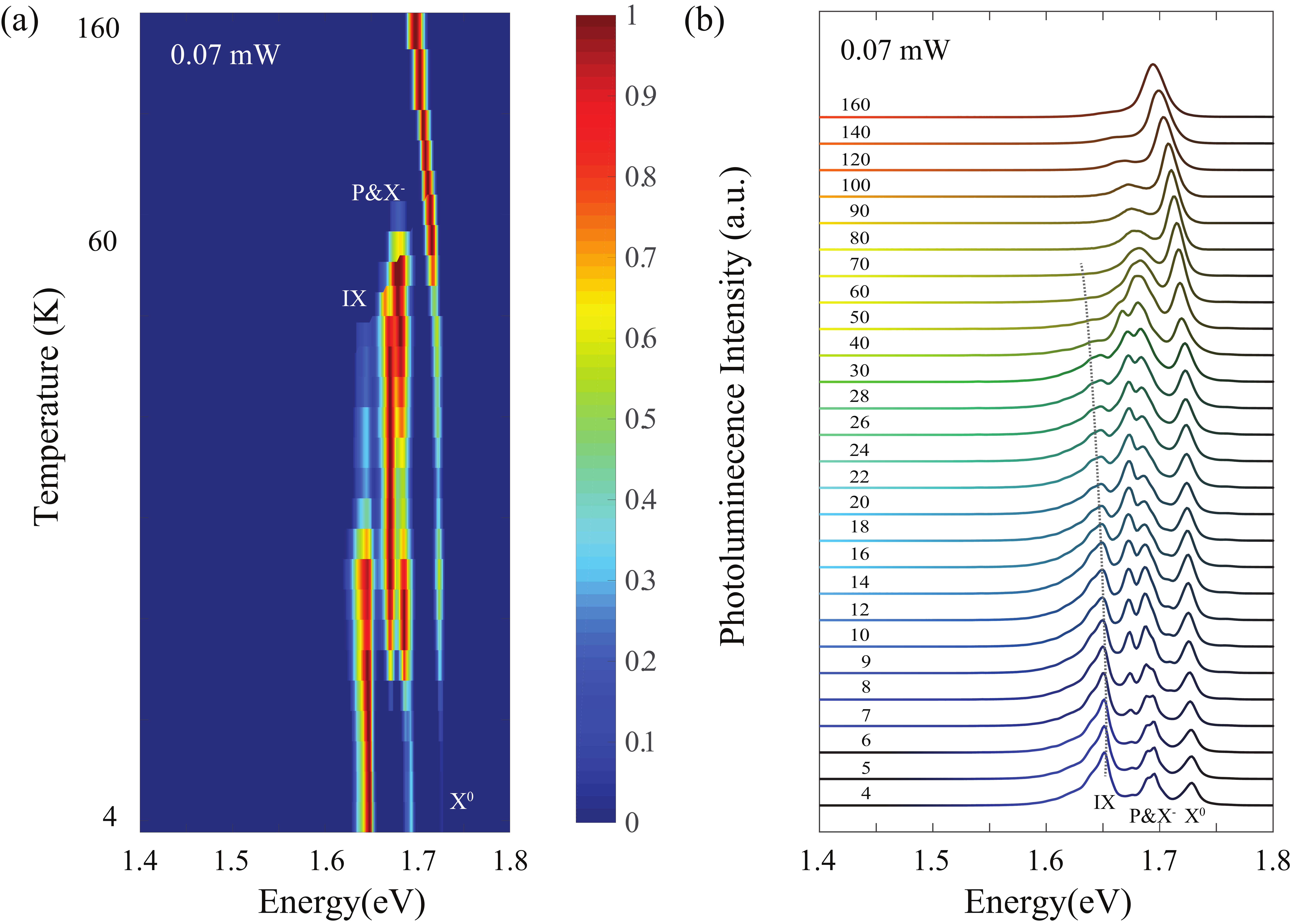}
	\caption{Two different visualizations of the temperature-dependent PL from Sample 2 created by pumping with a He-Ne laser with excitation power of 0.07 mW. The indirect exciton peak decreases as temperature increases and disappears around 60 K. The dashed gray line indicates the temperature-dependent bandgap shift.}
	\label{S:SI6}
\end{figure}

\begin{figure}[H]
	\centering
	\includegraphics[width=0.6\linewidth]{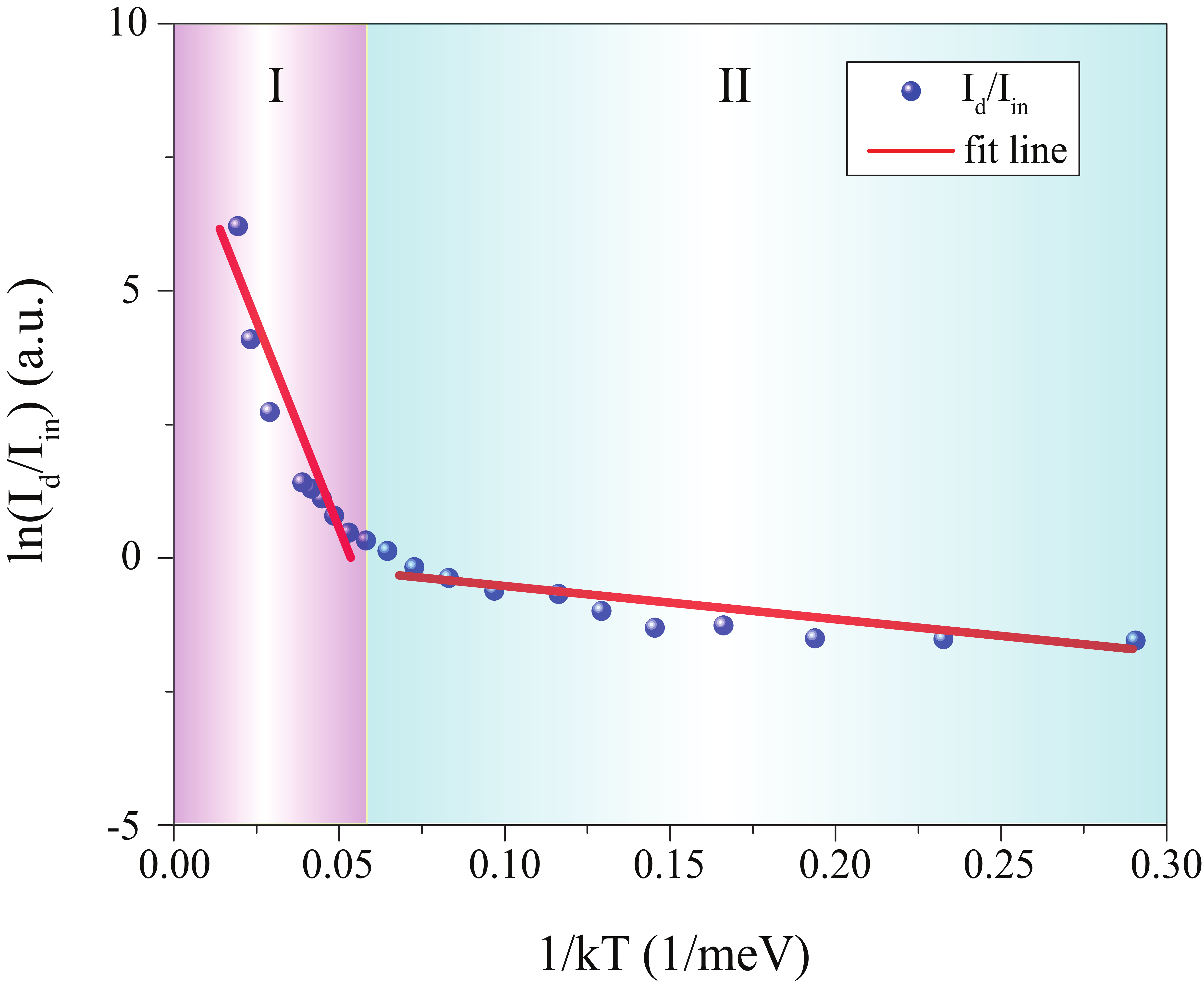}
	\caption{Ratio of integrated intensity of the direct and interlayer excitons in Sample 2 as a function of $1/kT$ plotted with a logarithmic axis, where $k$ is the Boltzmann constant. The colored lines give the results of the fits. The first regime and second regime are separated by different background colors, and correspond to temperatures above and below 22 K. The fitted slope in regime I and II are  -103$\pm$18meV and  -5$\pm$1meV respectively. }
	\label{S:SI7}
\end{figure}

\begin{figure}[H]
	\centering
	\includegraphics[width=0.9\linewidth]{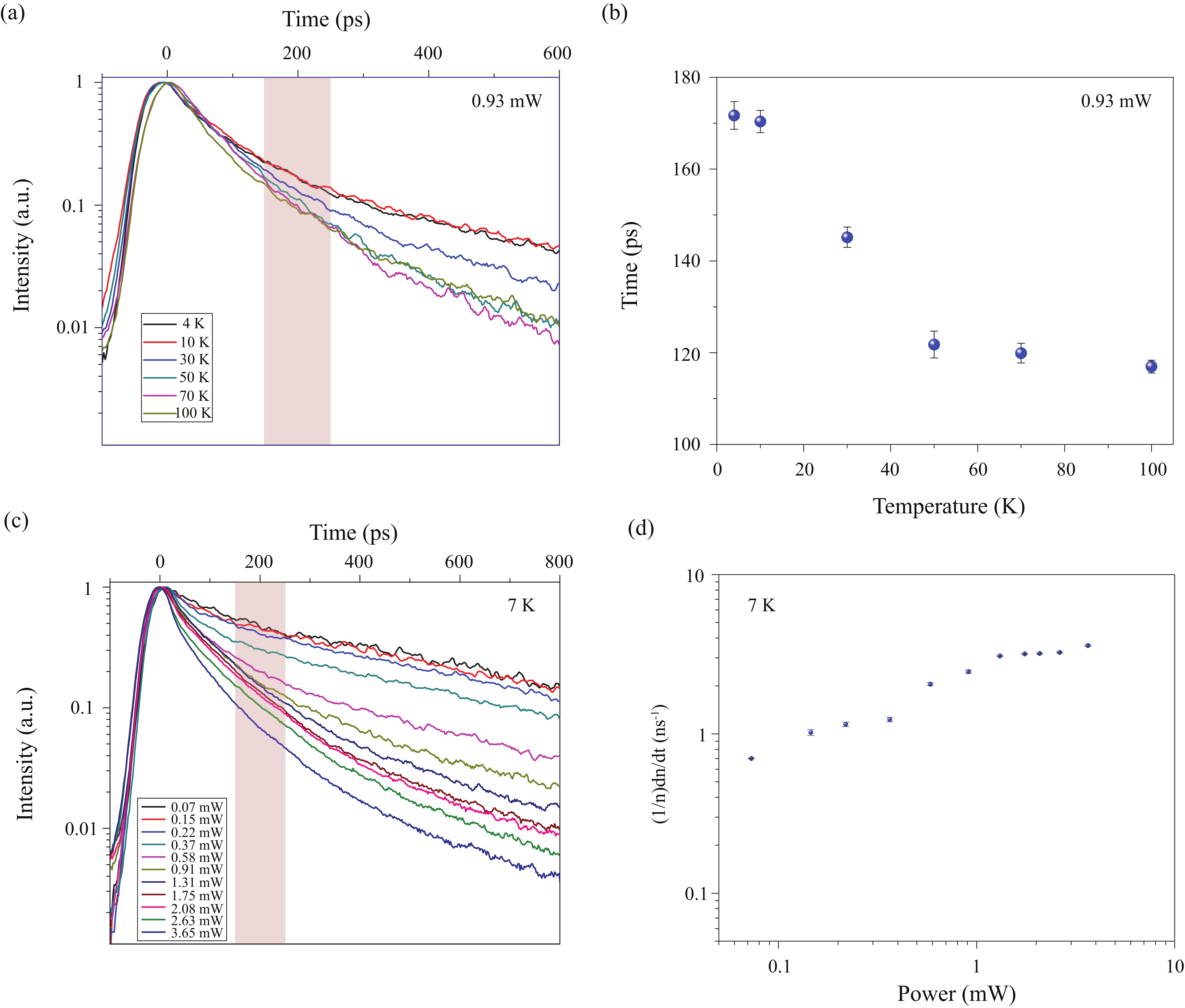}
	\caption{(a) (b) Temperature dependent time-resolved photoluminescence image from Sample 2, measured by streak camera. The interlayer exciton (1.65 eV) shows a lifetime of $\sim$575 ps under excitation power of 0.93 mW at 4 K by fitting the tail of the spectrum in S5 (a) at late times. It is hard to resolve the interlayer excitons emission when the temperature is higher than 80 K. (c) Power dependent time-resolved PL measured at 7 K. (d) Decay rate as a function of the pump power plotted on a log scale measured at 7 K. Decay rate increases significantly when the power above 0.5 mW. (b) and (d) show the density dynamic in the intermediate time window of 150-250 ps.}
	\label{S:SI8}
\end{figure}